\documentclass[pre,aps,reprint,superscriptaddress]{revtex4-1}
\usepackage{graphics}
\usepackage{graphicx}
\usepackage{amsmath}
\usepackage{float}
\usepackage{xcolor}
\usepackage{ulem}

\begin{document}

\title{Criticality in sheared, disordered solids. II. Correlations in avalanche dynamics} 

\author{Joel T. Clemmer}
\affiliation{Sandia National Laboratories, Albuquerque, New Mexico 87123, USA}
\author{K. Michael Salerno}
\affiliation{Army Research Lab, Aberdeen, Maryland 21005, USA}
\author{Mark O. Robbins}
\affiliation{Department of Physics and Astronomy, Johns Hopkins University, Baltimore, Maryland 21218, USA}
\date{\today}

\begin{abstract}

Disordered solids respond to quasistatic shear with intermittent avalanches of plastic activity, an example of the crackling noise observed in many nonequilibrium critical systems.
The temporal power spectrum of activity within disordered solids consists of three distinct domains: a novel power-law rise with frequency at low frequencies indicating anticorrelation, white-noise at intermediate frequencies, and a power-law decay at high frequencies.
As the strain rate increases, the white-noise regime shrinks and ultimately disappears as the finite strain rate restricts the maximum size of an avalanche.
A new strain-rate- and system-size-dependent scaling theory is derived for power spectra in both the quasistatic and finite-strain-rate regimes.
This theory is validated using data from overdamped two- and three-dimensional molecular dynamics simulations.
We identify important exponents in the yielding transition including the dynamic exponent $z$ which relates the size of an avalanche to its duration, the fractal dimension of avalanches, and the exponent characterizing the divergence in correlations with strain rate.
Results are related to temporal correlations within a single avalanche and between multiple avalanches.

\end{abstract}

\maketitle

\section{Introduction}

Power-law distributed bursts of motion, or avalanches, have been identified in a diverse range of driven disordered systems.
Such behavior has been termed crackling noise \cite{Sethna2001, Salje2014} and its origin is often linked to nonequilibrium criticality.
In this paper, we focus on the onset of flow in yield stress materials such as granular packings \cite{Miller1996, Hayman2011}, foams \cite{Park1994}, bubble rafts \cite{Durian1995, Dennin2004}, and emulsions \cite{Mason1996}. However, general results can be extended to other cases such as the depinning of elastic interfaces \cite{Fisher1998,Kardar1997} which also includes magnetic domain wall motion \cite{Cote1991, Perkovic1995,Durin2002}, fluid invasion in porous media \cite{Stokes1988, Martys1991, Moura2017}, and crack front propagation \cite{Gao1989, Ramanathan1997, Maloy2006}.
Yield stress materials only flow if driven at a stress above a critical threshold $\sigma_c$.
The transition from a jammed to flowing state at $\sigma_c$ is known as the yielding transition.
Infinitesimally near but above $\sigma_c$, the system flows quasistatically and flow is characterized by sporadic bursts of particle rearrangement or avalanches.
The number of avalanches is proportional to the energy released by the avalanche raised to the exponent $-\tau$.
For finite systems in the quasistatic (QS) regime, this distribution is truncated at a cutoff that scales as a power of linear system size $L$ with an exponent $\alpha$, sometimes labeled $d_f$. 

To fundamentally understand the macroscopic response of yield stress materials, one must characterize the dynamics of avalanches.
Avalanches are not instantaneous events, but evolve in time with a duration that grows as a power of the linear size of an avalanche with an exponent $z$, the dynamic exponent.
Temporal correlations in avalanche activity, quantified by the kinetic energy, can be identified by calculating power spectra of these signals.
In the QS regime, the noise spectra consist of distinct regimes of power-law scaling with frequency.
The cutoff frequencies and exponents of these regimes are fundamentally tied to the critical statistics of avalanches.
For instance on timescales shorter than the duration of the largest avalanche, spectra decay as a power of increasing frequency with an exponent near unity, sometimes referred to as 1/f-like noise.
This exponent depends on the value of the critical exponents $\alpha$, $\tau$, and $z$ \cite{Kuntz2000}.

As the stress $\sigma$ increases, the strain rate grows as a power of the distance from $\sigma_c$, $\dot{\epsilon} \sim (\sigma - \sigma_c)^{\beta}$. 
The increased strain rate restricts the maximum growth of avalanches and a finite correlation length $\xi$ emerges which scales as $\xi \sim |\sigma -\sigma_c|^{-\nu}$.
If $\xi < L$, the size of the largest avalanches is no longer governed by the system size $L,$ but rather by $\xi$.
This designates the finite strain rate (FSR) regime and leads to fundamental changes in spectra of kinetic energy.

In this work, we derive a new system-size and strain-rate dependent scaling theory for the noise spectra in the yielding transition.
This theory includes a description of the scaling of the magnitude of the power spectrum and the bounding frequencies of different power-law regimes, including a newly identified low-frequency regime.
We also characterize the evolution of power spectra with increasing rate as the system transitions from the QS to the FSR regime.

Molecular dynamics (MD) simulations of sheared disordered solids are used to test these relations and measure several critical exponents of the yielding transition including the dynamic exponent $z$ in two and three dimensions.
The dynamic exponent has been measured previously in elastoplastic models (EPMs) and found to be less than 1, contrary to physical restrictions \cite{Lin2014, Liu2016, Lin2018, Ferrero2019}. 
In contrast, the avalanche dynamics that emerge from our MD simulations obey $z > 1$, and represent distinct dynamical critical behavior.
We also measure the exponents $\alpha$ and the ratio $\beta/\nu$ by collapsing spectra obtained for different system sizes and strain rates.
Finally, our simulations also reveal new anticorrelations between large avalanches.
Correlations between avalanches have been seen before in systems such as sandpiles \cite{Hwa1992, Kutnjak-Urbanc1996} and earthquakes, where aftershocks empirically follow Omori's law \cite{Omori1894}.

This paper is divided into the following sections.
The model and details of the simulations are summarized in Sec. \ref{sec:method}.
In Sec. \ref{sec:yield_individual} we review QS results and describe the power spectra of individual avalanches.
These results are then used to derive the total power spectrum of the system in Sec. \ref{sec:yield_s(w)} and finite-size scaling relations are proposed and tested.
We then move to the FSR regime in Sec. \ref{sec:yield_s(w)_FSR} and derive a finite-rate scaling of power spectra and collapse FSR data onto a single curve. 
Lastly in Sec. \ref{sec:yield_s(w)_size}, we examine the crossover between the QS and FSR regimes.
The results in the paper are summarized in Sec. \ref{sec:yield_summary}.

\section{Method}
\label{sec:method}

Molecular dynamics (MD) simulations were run in two and three dimensions using LAMMPS \cite{Plimpton1995}.
Here we provide a brief summary of the methodology. 
Further details on the model, system preparation, and simulations can be found in the sibling paper \cite{Clemmer2021}. 
Both papers use data from the same set of simulations.

Systems consist of two types of particles, $A$ and $B$, with a number ratio of $N_A/N_B = (1+\sqrt{5})/4$.
Particles of type $i$ and $j$ interact with a Lennard-Jones (LJ) interaction with an energy scale of $u_{ij}$ and a length scale of $a_{ij}$. Interactions are smoothed to zero force and energy between a distance of $1.2 a_{ij}$ and a cutoff of $1.5 a_{ij}$. 
Interaction lengths of $a_{BB} = 0.6 a_{AA}$ and $a_{BA} = 0.8 a_{AA}$ and energy scales of $u_{BB} = u_{AA}$ and $u_{AB} = 2 u_{AA}$ are chosen to prevent crystallization during flow.
The fundamental units of length and energy are set to $a = a_{AA}$ and $u = u_{AA}$, respectively.
Particles of both types have mass $m$.
All subsequent values in this paper are scaled by the appropriate combination of these constants and are dimensionless.

Systems were prepared by randomly placing particles in a square or cubic box.
Particles were shifted using a soft cosine potential to remove overlap before switching to the LJ interaction.
The system was then expanded to a final density of $\rho = 1.7$ and $1.4$ in 2D and 3D, respectively. 
Rounded to the nearest integer, systems with side lengths of $L=20$, 41, 81, or 163 in 3D and $55$, 110, 219, 438, 877, or 1753 were studied.

After preparation, systems underwent pure shear deformation.
The $x$ dimension of the box was expanded at a constant true strain rate of $\dot{\epsilon}$ while the other dimension(s) were contracted to preserve volume.
To reach steady-state flow at large strains, we used Kraynik-Reinelt (KR) boundary conditions in 2D \cite{Kraynik1992} and generalized KR boundary conditions in 3D \cite{Hunt2016}.
Deformation was enforced by affinely remapping particles positions within the simulation box.

The stress tensor $\sigma_{\alpha \beta}$ was calculated with contributions from both the virial stress and kinetic energy associated with the nonaffine velocity of particles \cite{Allen1989}.
A shear stress was defined as $\sigma = (\sigma_{xx}-\sigma_{yy})/2$ in 2D and $\sigma = (2\sigma_{xx} - \sigma_{yy} - \sigma_{zz})/4$ in 3D.
Energy was removed using a viscous damping force equal to $\vec{F}_{i,\mathrm{damp}} = - \frac{1}{2} \Gamma \vec{v}_{i,\mathrm{na}}$ where $\vec{v}_\mathrm{na}$ is the nonaffine velocity of particles and $\Gamma = 4.0$ is large enough to overdamp the system \cite{Salerno2012, Salerno2013}. 
On average, the power removed from the system is
\begin{equation}
P = \sum_i \frac{1}{2} \Gamma \vec{v}_{i,\mathrm{na}}^2 = \Gamma \langle K
\label{eq:power}
\rangle
\end{equation}
where $K$ is the extensive kinetic energy of the system.

\section{QS avalanche statistics}
\label{sec:yield_individual}

We begin with a brief review of avalanches and scaling relations in the QS regime.
A more complete review can be found in the sibling paper \cite{Clemmer2021}.
Each avalanche is associated with a drop in the stress $\delta \sigma_I$ and energy $E_I$ of the system.
For sufficiently large avalanches, $\delta \sigma_I$ is proportional to $E_I$: 
\begin{equation}
E_I = L^d \langle \sigma \rangle \delta \sigma_I/ 4\mu
\label{eq:aval_e_s}
\end{equation}
where $\mu$ is the shear modulus and $\langle \sigma \rangle$ is the average stress of the system \cite{Salerno2012, Salerno2013}.
Each avalanche also produces a burst of kinetic energy with a profile $k_I(t)$.
From Eq. \eqref{eq:power}, energy is dissipated at a rate proportional to the kinetic energy implying
\begin{equation}
E_I \sim \Gamma \int dt k_I(t) \ \ .
\label{eq:aval_e_k}
\end{equation}
Equations \eqref{eq:aval_e_s} and \eqref{eq:aval_e_k} suggest there is an equivalence between $K$ and $L^d d \sigma/dt$.
This equivalence is verified in Appendix \ref{sec:ke_v_s}.

\begin{figure}
\begin{center}
	\includegraphics[width=0.45\textwidth]{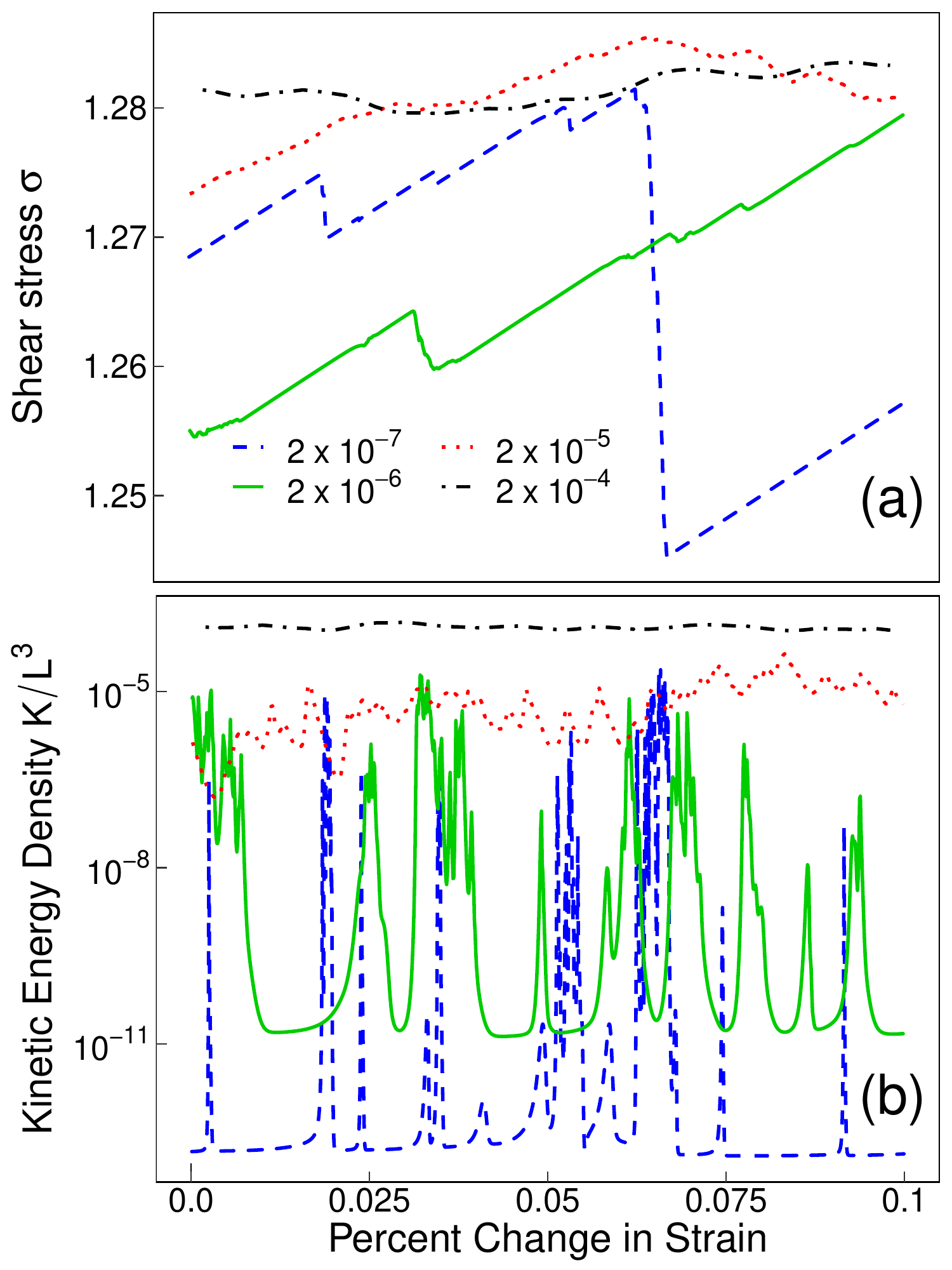}
	\caption{
	A snapshot of the (a) shear stress and (b) kinetic energy per particle plotted over an interval of 0.1\% strain in steady state flow. 
	Data is from a single 3D system strained at the indicated rates with $L = 40$. Similar trends are seen in 2D. 
	} \label{steady_flow}
	\end{center}
\end{figure}

Typical curves of the shear stress and kinetic energy versus strain in steady state flow are plotted in Fig. \ref{steady_flow} for a 3D system with $L = 40$. 
For now, we focus on data for $\dot{\epsilon} = 2\times 10^{-7}$ which is characteristic of the QS regime.
In Fig. \ref{steady_flow}(a), the stress grows linearly with strain until the system becomes unstable and an avalanche nucleates leading to a rapid drop in stress.
Similar behavior is seen in the potential energy.
In Fig. \ref{steady_flow}(b), each avalanche also corresponds to a spike in the kinetic energy.
Note that the smallest avalanches have stress drops less than the width of the line and are only visible in the trace of kinetic energy.

The avalanche rate distribution $R_{QS}(E,L)$ is defined as the average number of avalanches nucleated in a unit strain with a magnitude of $E$ in a system of size $L$ in the QS limit. 
The rate has a functional form of
\begin{equation}
	R_{QS}(E,L) \sim L^\gamma E^{-\tau} f(E/L^\alpha)
\label{eq:aval_rate}
\end{equation} 
where $\alpha$, $\tau$, and $\gamma$ are critical exponents and $f(x)$ is constant for $x \ll 1$ and rapidly goes to zero for $x > 1$ or $E > E_\mathrm{max} \sim L^\alpha$.
QS simulations have found the rate distribution of $\delta \sigma_I$ is characterized by the same exponents \cite{Salerno2012,Salerno2013}.

Energy conservation requires that the cumulative energy released by all avalanches equal the work done to the system during shear.
Therefore the integral of $E R_{QS}(E,L)$ is proportional to $\langle \sigma \rangle L^d \dot{\epsilon}$.
From this observation, one can derive
\begin{equation}
	\gamma+(2-\tau)\alpha=d
	\label{eq:gamma}
\end{equation}
as argued in Refs. \cite{Salerno2012, Salerno2013, Clemmer2021}.
Notably, $\gamma < d$ implying the number of avalanches grows sub-extensively with system size. 
This is a significant difference from from depinning where the number of avalanches is extensive \cite{Martys1991a,Clemmer2019}.

The evolution of  $K$ and $\sigma$ with strain in Fig. \ref{steady_flow} contains additional information. 
In particular, the signals encode the dynamical structure of individual avalanches that can be revealed by calculating the temporal power spectra of $K$ or $\sigma$.
As noted above, both $K$ and $L^d d \sigma/dt$ are proportional to the rate of energy dissipation and thus contain similar information.
The numerical factors in the definition of the power spectrum and the method of numerical evaluation are described in Appendix A. 
In the main text, we focus only on the power spectra of $K$ and note that similar scaling behavior is seen for spectra of $\sigma$.

As shown in Appendix A, if there are no correlations between avalanches, the total noise power can be written as an integral over the noise power, $S_k(E,\omega)$, of avalanches of energy $E$.
In the QS regime,
\begin{equation}
	S_K(\omega) = \int dE \  \dot{\epsilon} \  R_{QS} (E,L) \  S_k(E,\omega) \ \ ,
\label{eq:S_Kint}
\end{equation}
where $R_{QS} \sim L^\gamma E^{-\tau}$ for $E< E_\mathrm{max}\sim L^\alpha$ from Eq. \eqref{eq:aval_rate}.
Further progress requires information about the scaling of $S_k(E,\omega)$.
From Eq. \eqref{eq:aval_e_k}, the integral of the kinetic energy in an avalanche is $E/\Gamma$, and thus $S_k(E,0)=E^2/\Gamma^2$.
The power spectrum will drop below this low-rate limit when $\omega$ exceeds
the lowest frequency in an avalanche $\omega_\mathrm{min}(E)$.
The duration $T$ of an avalanche scales as a power of its linear extent $\ell$ with an exponent $z$ known as the dynamic exponent: 
\begin{equation}
T(E) \sim \ell^z \sim E^{z/\alpha} \ \ .
\label{eq:dynamic_exp}
\end{equation}
Given this, 
\begin{equation}
\omega_\mathrm{min}(E)  \equiv 2\pi/T(E) \sim E^{-z/\alpha} \ \ .
\label{eq:wmin}
\end{equation}

Avalanches are observed to have bursts of activity that lead to power at a wide range of frequencies greater than $\omega_{\mathrm{min}}(E)$.
Thus, we assume a power-law power spectrum up to a maximum frequency $\omega_{\mathrm{max}}$ that is independent of $E$ and must be comparable to the speed of individual atomic rearrangements.
These limiting scaling properties are satisfied by:
\begin{equation}
S_k(E,\omega) = \frac{E^2}{\Gamma^2} \left( 1+ b \omega^{\alpha/z} E\right)^{-q}
		\label{eq:pi}
\end{equation}
where $b$ is some constant, $q$ is a new exponent, and
the high frequency power spectrum $\sim\omega^{-q \alpha/z}$ up to $\omega_{\mathrm{max}}$.
Note that the functional form of the crossover to this power-law behavior may be different than assumed in Eq. \eqref{eq:pi}, but neither this nor the value of $b$ affect the scaling relations derived below.

To confirm Eq. \eqref{eq:pi} we calculated the average power spectrum for avalanches in 2D simulations using the algorithm of Ref. \cite{Salerno2013} where strain was stopped during the avalanche.
The power spectra were averaged over avalanches within a range of energies centered on 2.4 to 38.57. The smallest avalanches are just at the start of the scaling regime for the distribution of avalanches, and the largest are small enough that they are not limited by system size for $L=438$ \cite{Salerno2013}.

\begin{figure}
\begin{center}
	\includegraphics[width=0.45\textwidth]{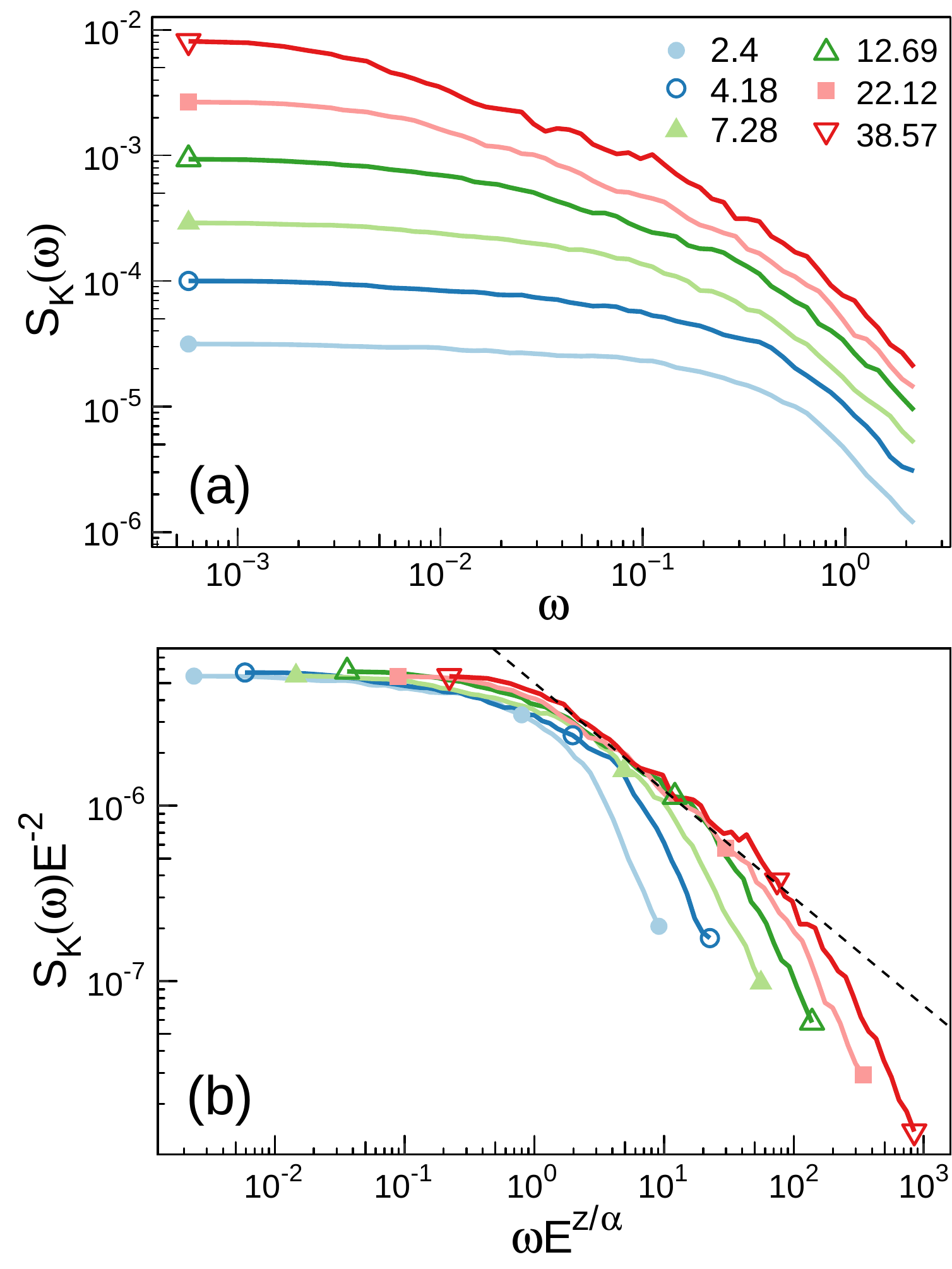}
	\caption{(a) Average power spectra $S_k(E,\omega)$ for avalanches in a range of energies near the value indicated in the legend for 2D simulations with $L=438$ using the algorithm of Ref. \cite{Salerno2013}.
	(b) Same data rescaled according to Eq. \eqref{eq:pi} 
	with $z/\alpha=1.63$. Data collapse onto a common curve for $\omega$ below 0.2, which is indicated by a symbol on each curve.
The dashed line decays as a power law at high frequencies with an exponent consistent with the value of $q \alpha/z=0.61$.
} 
	\label{fig:avalanchepower}
	\end{center}
\end{figure}

Figure \ref{fig:avalanchepower}(a) shows raw data for each range of energies.
We see an increase in the rate of decay of the power spectrum after $\omega =0.2$, which is indicated by a symbol for each curve in Fig. \ref{fig:avalanchepower}(b).
The drop is significant by $\omega =0.5$ and we take this as an estimate
of $\omega_{\mathrm{max}}$ for 2D in the following.
Figure \ref{fig:avalanchepower}(b) shows the same power spectra rescaled by $E^2$ against $\omega E^{z/\alpha}$ for $\omega < \omega_\mathrm{max}$.
As predicted by Eq. \eqref{eq:pi}, this collapses the data quite well, particularly for the larger energies, using a value of $z/\alpha = 1.63$ as determined by the following analysis of constant strain rate results.
There is a power-law decay in the spectrum at high frequencies over almost 2 decades.
Moreover the decay is consistent with the exponent $q \alpha/z =0.61$ (dashed line) which was also determined by subsequent analysis in the following section.

\section{Frequency dependent noise power of $K$ in the QS regime }
\label{sec:yield_s(w)}

Having described the statistics and noise power of individual avalanches, we now turn back to the total power spectrum obtained from Eq. \eqref{eq:S_Kint}.
Integrating contributions from all avalanche sizes using Eq. \eqref{eq:aval_rate} yields a total power of:
\begin{equation}
	S_K(\omega)\sim \dot{\epsilon} L^\gamma \int^{E_{\mathrm{max}}} dE 
	E^{2-\tau} \left( 1+b \omega^{\alpha/z} E\right)^{-q} \ \ ,
\label{eq:S_Kint2}
\end{equation}
where numerical prefactors such as $\Gamma$ are dropped here and below.
Changing variables to $v \equiv b \omega^{\alpha/z} E$ one can rewrite the integral as:
\begin{equation}
	S_K(\omega) \sim \dot{\epsilon} \ L^{\gamma} \omega^{(\tau-3)\alpha/z} \int_0^{v_{\mathrm{max}}} \frac {dv \ v^{2-\tau}}{ (1+v)^{q}} \ \ .
\label{eq:Sint}
\end{equation}
Past work shows $2-\tau >0$ for this system \cite{Salerno2012,Salerno2013}.
This implies the final integral will be proportional to $ v_{\mathrm{max}}^{3-\tau}$ or $\omega^{(3-\tau)\alpha/z} E_\mathrm{max}^{3-\tau}$ for small $\omega$ where $v_{\mathrm{max}} \ll 1$.
In this small $\omega$ limit, the frequency cancels out and substituting for $E_\mathrm{max} = L^\alpha$ yields:
\begin{equation}
	S_K(\omega) \sim \dot{\epsilon} L^{\gamma+(3-\tau)\alpha}
	\sim \dot{\epsilon} L^{d + \alpha}
	\  \ ,
\label{eq:Sk_Plateau_qs}
\end{equation}
where the final equation follows from the scaling relation for $\gamma$ in Eq. \eqref{eq:gamma}.

For high frequencies, where $v_{\mathrm{max}}$ is large, the integral in Eq. \eqref{eq:Sint} depends on $q$. 
If $q> 3-\tau$, then the integral is dominated by $v<1$ and becomes a constant at
large $v_{\mathrm{max}}$. Instead, as in past work on spin systems \cite{Kuntz2000}, we find that $q < 3-\tau$ and the integral is dominated by large values of $v$, scaling as $v_{\mathrm{max}}^{3-\tau-q} \sim \omega^{(3-\tau-q)\alpha/z} L^{\alpha(3-\tau-q)}$.
Thus
\begin{align}
\begin{split}
	S_K(\omega) &\sim \dot{\epsilon} L^{\gamma+(3-\tau-q)\alpha} \omega^{-q \alpha/z} \\
	&\sim \dot{\epsilon} L^{d+(1-q) \alpha} \omega^{-q \alpha/z},
	\label{kscale}
\end{split}
\end{align}
for $\omega > \omega_{\textrm{min}}(E_\textrm{max}) \sim L^{-z}$ using the scaling relation for $\gamma$ from Eq. \eqref{eq:gamma}.

A constraint on $q$ can be derived using Eq. \eqref{kscale} and the scaling of the variance of $K$.
In the QS regime, the same sequence of avalanches occurs over a longer time interval as $\dot{\epsilon}$ decreases.
Therefore, the only change in $K(t)$ will be that the duration of quiescent periods between avalanches grows by a factor of $1/\dot{\epsilon}$. 
Therefore, integrals of $K$ or $K^2$ will be constant.
After normalizing by time, averages will scale as $\dot{\epsilon}$.
This argument also explains the trivial dependence on $\dot{\epsilon}$ in Eq. \eqref{eq:Sk_Plateau_qs}.
The variance $\Delta K^2 \equiv \langle K^2 \rangle - \langle K \rangle^2$ in the low rate limit will be dominated by $\langle K^2 \rangle$ as $\langle K \rangle^2$ will scale as $\dot{\epsilon}^2$.
Therefore, the integral of $S_K(\omega)$ which is equal to $\langle K^2 \rangle$ (Eq. \eqref{eq:meanK2v2}), is approximately $\Delta K^2$ (Eq. \eqref{eq:devK}).
If $q \alpha/z < 1$, then the integral of $S_K$ over $\omega$ will be dominated by large frequencies.
Given that
$\omega_\mathrm{max}$ is assumed to be independent of $L$, rate, and other factors, integrating Eq. \eqref{kscale} gives
\begin{equation}
	\Delta K^2 
	\sim \int \frac{d\omega}{2\pi} S_K(\omega) \sim \dot{\epsilon} L^{d+(1-q)\alpha} \ \ .
	\label{eq:devK2}
\end{equation} 
In Ref. \cite{Clemmer2019}, we found that $\Delta K$ scales as $L^{d/2}$ with an uncertainty of less than 0.1, implying $q=1$.

\begin{figure}
\begin{center}
	\includegraphics[width=0.45\textwidth]{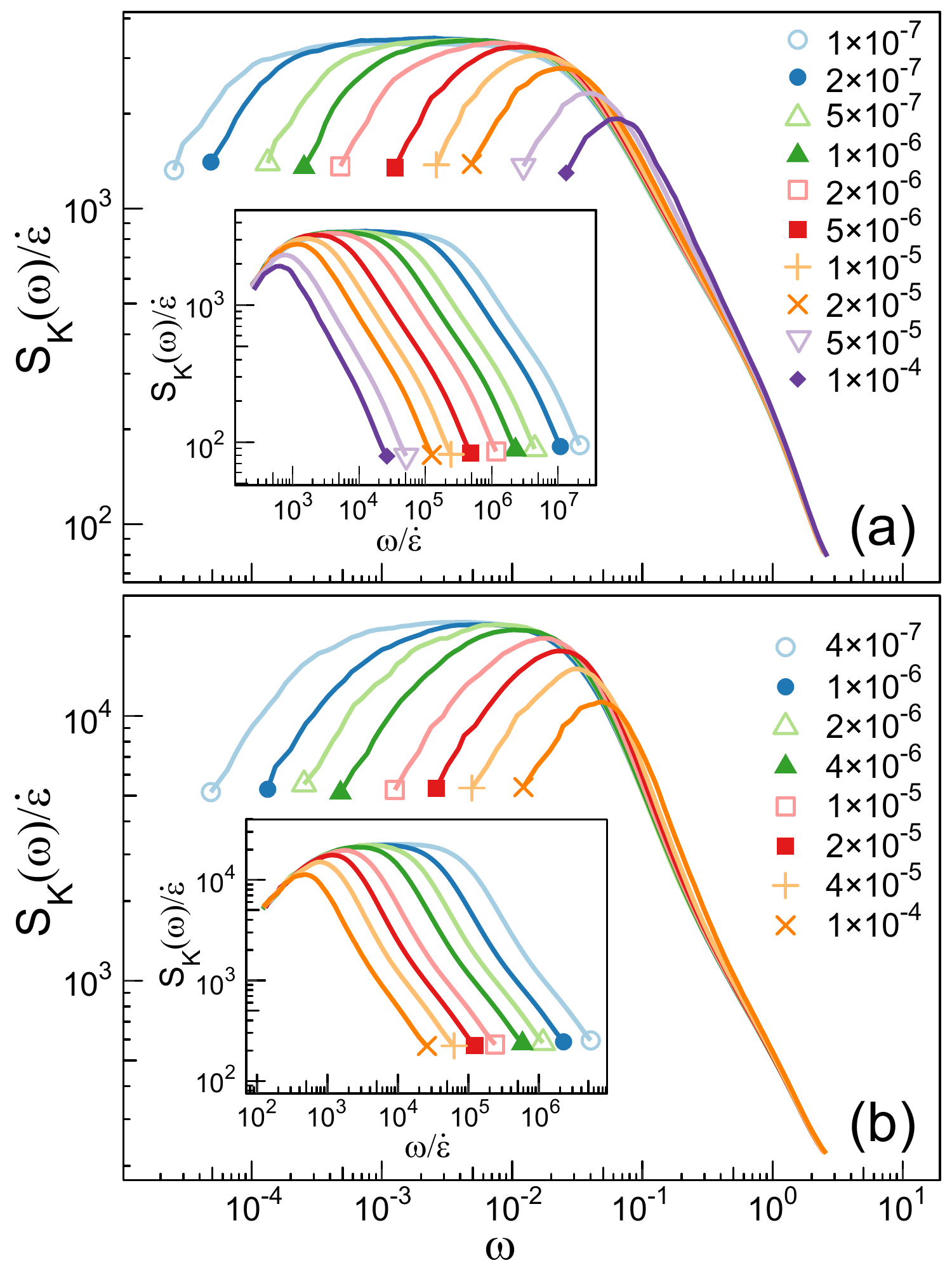}
	\caption{ Power spectra normalized by rate for the indicated rates in (a) 2D systems with $L=55$ and (b) 3D systems with $L=20$. The insets in each panel depict the same data after scaling the frequency by the strain rate to collapse the low frequency data.} 
	\label{fig:ratedep}
	\end{center}
\end{figure}

Figure \ref{fig:ratedep} illustrates the scaling of $S_K (\omega)$ with $\omega$ as a function of rate for small systems.
The low rate behavior is generally consistent with the relations derived above.
There is a pronounced plateau, equivalent to white noise, whose height is proportional to $\dot{\epsilon}$.
At higher frequencies, $S_K (\omega)$ decays as a power of $\omega$.
As noted above, this power-law scaling should only extend up to an
$\omega_\mathrm{max}$ that is comparable to the rates of atomic rearrangements and damping.
The data shows a change in scaling at $\omega > \omega_{\mathrm{max}} \sim 0.5$ in 2D, as seen also in Fig. \ref{fig:avalanchepower}. The value of $\omega_{\mathrm{max}}$ is approximately $2$ in 3D and is more evident when higher frequencies are included in plots.
These values of $\omega_{\mathrm{max}}$ are comparable to the frequency where phonons become important and where the damping force affects the response: $\omega > 2 \pi/\Gamma \sim 1.5$.

The main deviation from the above relations is at low $\omega$.
The plateau is preceded by a region where $S_K (\omega)$ rises as a power of $\omega$ with an exponent defined as $\eta$.
This power law extends up to a frequency $\omega_\mathrm{corr}(\dot{\epsilon}, L)$ that rises with rate. At high enough rates the plateau disappears and there is a single peak in $S_K (\omega)$ that moves to higher frequency with increasing rate and decreases in amplitude as the system enters the FSR regime.
The rise in $S_K (\omega)$ with increasing $\omega$ for $\omega < \omega_\mathrm{corr}$ implies a breakdown in the assumption that avalanches are decorrelated.
The insets to Fig. \ref{fig:ratedep} show that $\omega_\mathrm{corr}(\dot{\epsilon},L)$ scales linearly with rate, implying that correlations set in above the same strain interval for all rates.

A possible mechanism for correlations is that large avalanches lead to large energy and stress drops and thus make similar large events unlikely until strain has accumulated again.
The rate of avalanches per unit strain that span the system, $R_{\textrm{span}}$, can be calculated by integrating $R_{QS}(E,L)$ over energies of order $E_{\textrm{max}}$.
Taking a lower bound $c E_{\textrm{max}}$ where $c$ is a constant less than unity,
\begin{align}
\begin{split}
	R_\mathrm{span} 
	& \sim \int_{cE_{\textrm{max}}}^{E_{\textrm{max}}} dE \ R_{QS}(E,L)  \\
	& \sim L^\gamma \int_{cE_{\textrm{max}}}^{E_{\textrm{max}}} dE E^{-\tau} \\
	& \sim L^{\gamma+ \alpha(1-\tau)} \sim L^{d-\alpha}, \\
\end{split}
\end{align}
where the scaling with $L$ does not depend on $c$ and the final relation follows from Eq. \eqref{eq:gamma}.
Multiplying by $\dot{\epsilon}$ to convert $R_\mathrm{span}$ to a rate per unit time, gives the prediction
\begin{equation}
	\omega_\mathrm{corr}(\dot{\epsilon},L) \sim \dot{\epsilon} L^{d-\alpha} \ \ ,
	\label{eq:wcorr}
\end{equation}
which is consistent with the linear scaling with rate in Fig. \ref{fig:ratedep}.

\begin{figure*}
\begin{center}
	\includegraphics[width=0.9\textwidth]{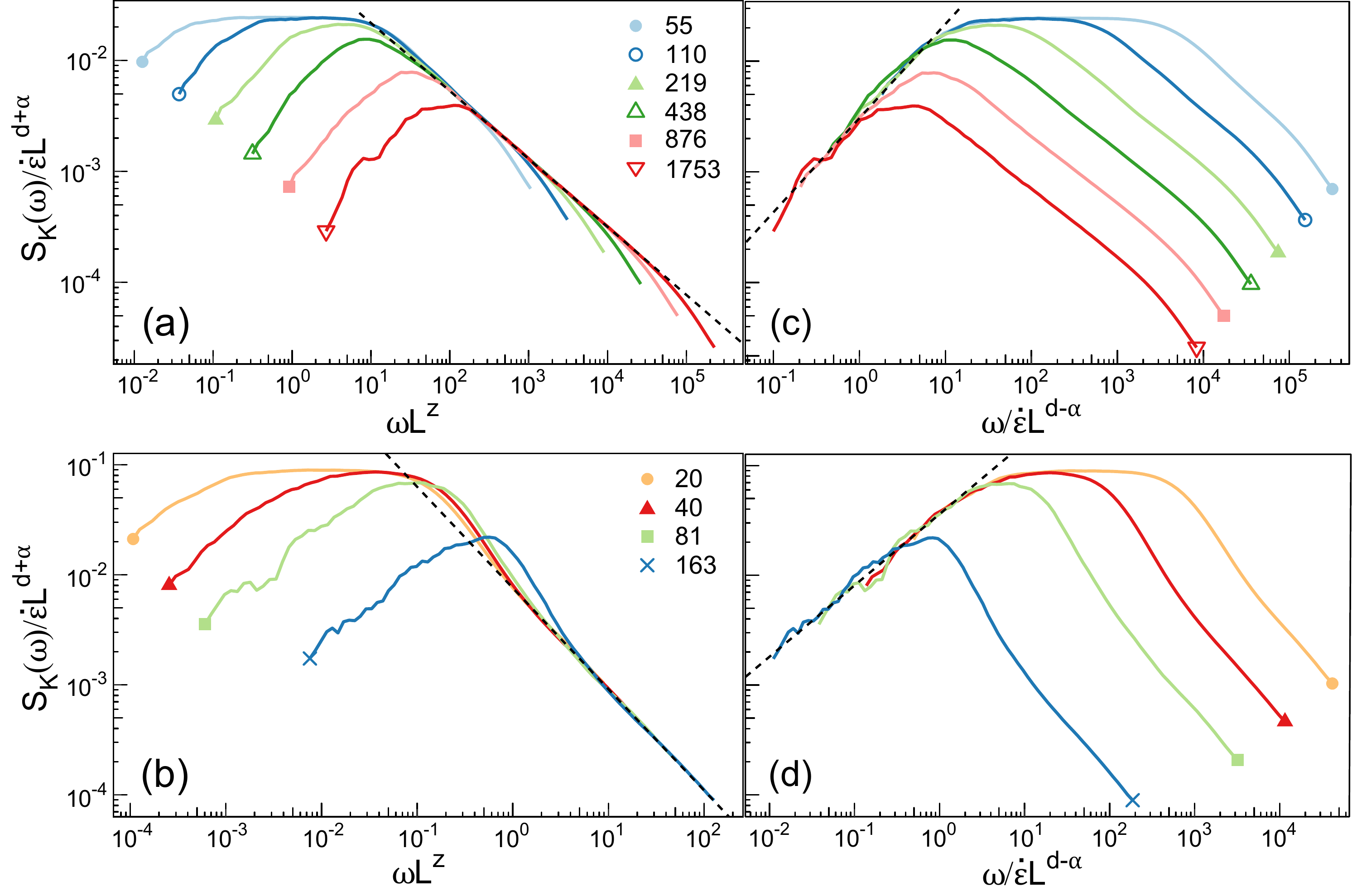}
	\caption{ Power spectra for the indicated $L$ for (a) 2D systems at $\dot{\epsilon}=10^{-7}$ and (b) 3D systems at $\dot{\epsilon}=2\times 10^{-7}$ for $L \leq 81$ and $\dot{\epsilon}=10^{-6}$ for $L = 163$.
	Curves are normalized by $ \dot{\epsilon} L^{\alpha+d}$ with $\alpha=0.95$ in 2D and 1.15 in 3D. The frequency is shifted by $L^z$ with $z=1.55$ in 2D and 1.25 in 3D.
	Panels (c) and (d) show 
	the same data scaled to collapse the low frequency regime. The collapse shows that the plateau begins at a strain that decreases with increasing $L$ as $L^{-(d-\alpha)}$.
	Dashed lines represent power laws with exponents $-q \alpha/z = -0.61$ (a), $-q \alpha/z = -0.92$ (b), $\eta = 0.85$ (c), and $\eta = 0.65$ (d).
} 
	\label{fig:QSscale}
	\end{center}
\end{figure*}

Figure \ref{fig:QSscale} shows how the QS power spectra change with $L$ for the lowest rates studied. 
These rates are not low enough to produce a plateau for the largest system sizes, implying that the QS regime has not been reached for all $L$. 
In the sibling paper, we looked at the finite-size scaling of the average shear stress as a function of strain rate \cite{Clemmer2021}.
We found that the stress deviates from an asymptotic power law for $\dot{\epsilon} L^{\beta/\nu}$ less than about 2 and approaches a limiting constant less than $\sigma_c$ at rates an order of magnitude lower.
Therefore, only systems in this asymptotic regime ($L \leq 219$ in 2D and $40$ in 3D) indicate a clear plateau in $S_K (\omega)$ and thus fully exhibit QS scaling.

The collapse of data for different $L$ in Figs. \ref{fig:QSscale}(a) and (b)
confirm that the height of the QS plateaus scales as $L^{d+\alpha}$, as predicted by Eq. \eqref{eq:Sk_Plateau_qs}, and the plateau ends at the low frequency end of power-law scaling $\omega_{\textrm{min}}(E_{\textrm{max}}) \sim L^z$, as predicted by Eq. \eqref{kscale}.
The scaling of the data is consistent with values of $\alpha = 0.95 \pm 0.1$ ($1.15 \pm 0.1$) and $z = 1.55 \pm 0.1$ ($1.25 \pm 0.1$) in 2D (3D). 
These measured values of $\alpha$ are consistent with measurements from Refs. \cite{Salerno2012} and \cite{Salerno2013}. Note that the large error bars are due to the limited amount of data which exhibits a clear plateau. Estimates of $z$ will be refined in the subsequent section.
Even systems without a plateau collapse onto a common high frequency power law,
only deviating in a more rapid drop at $\omega > \omega_{\mathrm{max}} \sim 0.5$.
The exponent of this power-law drop is consistent with values of $q\alpha/z$ determined below.

Figures \ref{fig:QSscale}(c) and (d) show that
curves for different $L$ at the same $\dot{\epsilon}$ collapse at low frequencies when $\omega$ is multiplied by $L^{d-\alpha}$.
This is true even for systems that are too large to show a plateau at this rate.
As noted above, spectra for the same $L$ at different rates show $\omega_\mathrm{corr}(\dot{\epsilon},L)$ is proportional to strain rate, implying correlations start at a rate independent strain. 
These two observations confirm that Eq. \eqref{eq:wcorr} describes the
scaling of $\omega_\mathrm{corr}$
and that the strain where correlations become important
decreases with increasing $L$ as $L^{-(d-\alpha)}$ in the QS regime.
Since the collapse in this plot only depends on $\alpha$ for both scaling in horizontal and lateral directions, it provides the tightest bounds on $\alpha$.
Combining data for all $L$ and rates we find $\alpha=0.95 \pm 0.05$ in 2D and $\alpha=1.15 \pm 0.05$ in 3D.

\section{Frequency dependent noise power of $K$ in the FSR regime }
\label{sec:yield_s(w)_FSR}

Having described the scaling of the noise power in the QS regime, we now turn our focus to the FSR regime. 
As the strain rate increases, energy needs to be dissipated at a quicker rate requiring additional avalanches nucleate before previous events complete.
This can be seen in Fig. \ref{steady_flow}.
At a higher rate of $2\times 10^{-6}$, the minimum value of $K$ has risen and there is no clear demarcation between peaks.
As the rate continues to rise, peaks disappear and there is constant activity: at least one event is always evolving in the system.
This temporal overlap between avalanches introduces a finite correlation length $\xi$ in the system which diverges at the critical stress with an exponent $\nu$:
\begin{equation}
\xi \sim | \sigma-\sigma_c|^{-\nu} \ \ .
\label{eq:xi}
\end{equation}
This correlation length represents the maximum extent of cooperative particle motion or the maximum linear size of an avalanche.
Above $\sigma_c$, the strain rate grows as a power of the distance from the critical stress:
\begin{equation}
\dot{\epsilon} \sim (\sigma-\sigma_c)^{\beta}
\end{equation}
as seen in Ref. \cite{Clemmer2021} implying 
\begin{equation}
\xi \sim \dot{\epsilon}^{-\nu/\beta} \ \ .
\label{eq:xi_rate}
\end{equation}
This relation implies that the crossover between the FSR and QS regimes, $\xi = L$, occurs at a rate of $\dot{\epsilon}_{QS} \sim L^{-\beta/\nu}$.

The scaling of $\omega_\mathrm{corr}(\dot{\epsilon},L)$ and $\omega_\mathrm{min}(E_\mathrm{max})$ in the previous section can be used to predict the start of the FSR regime.
The plateau will disappear when the strain rate is high enough that $\omega_\mathrm{corr} (\dot{\epsilon},L) > \omega_\mathrm{min}(E_{\mathrm{max}})$. Using Eqs. \eqref{eq:wmin} and \eqref{eq:wcorr}, this inequality can be rewritten as
\begin{align}
\begin{split}
	\dot{\epsilon }_{QS} L^{d-\alpha} \gtrsim & L^{-z} \\
	\dot{\epsilon }_{QS} \gtrsim &L^{-(d+z-\alpha)}  \ \ .
\end{split}
\end{align}
Using Eq. \eqref{eq:xi_rate}, we can therefore bound $\beta/\nu$:
\begin{equation}
\beta/\nu \le d+z-\alpha \ \ .
\end{equation}
The opposite bound was derived in the sibling paper \cite{Clemmer2021} implying there is actually an equality as proposed using different arguments in Ref. \cite{Lin2014}. 
Values of $\alpha$, $z$, and $\beta/\nu$ (measured below and in Ref. \cite{Clemmer2021}) are consistent with this relation.
The collapse of power spectra without plateaus at high frequencies in Fig. \ref{fig:ratedep} and low frequencies in Figs. \ref{fig:QSscale}(c) and (d) supports this prediction.

Expressions for the noise power in the FSR regime can be derived using the same approach as in the last section,
but remembering that the assumption that avalanches are not correlated will break down at low $\omega$.
The maximum avalanche is now cut off by $\xi$ instead of $L$ and its energy scales as $E_{\mathrm{max}} \sim \xi^\alpha$.
There will be of order $(L/\xi)^d$ independent spatial regions, giving a rate distribution $R \sim (L/\xi)^d \xi^\gamma E^{-\tau}$. Following the analysis in Sec. \ref{sec:yield_s(w)} we find
\begin{equation}
	S_K(\omega) \sim \dot{\epsilon} L^{d} \xi^{\gamma-d} \omega^{(\tau-3)\alpha/z} \int_0^{v_{\mathrm{max}}} \frac {dv \ v^{2-\tau}}{ (1+v)^{q}} \ \ ,
\label{eq:S_KFSR1}
\end{equation}
where again $v \sim \omega^{\alpha/z} E$.
As above, the low frequency limit corresponds to $v_\mathrm{max} \ll 1$, which now implies
$\xi^\alpha \omega^{\alpha/z} \ll 1$ or
$\omega < \xi^{-z} \sim \dot{\epsilon}^{z\nu/\beta}$.
In this limit, the integral over $v$ is again $\sim v_{\mathrm{max}}^{3-\tau}$ and
the frequency dependence cancels. One finds:
\begin{equation}
	S_K(\omega) \sim \dot{\epsilon} L^d \xi^{\gamma-d +(3-\tau)\alpha} 
	\sim \dot{\epsilon} L^d \xi^\alpha \sim L^d \dot{\epsilon}^{1- \nu \alpha/\beta} \ \ , 
	\label{eq:S_KFSR2}
\end{equation}
given the scaling relation for $\gamma$ in Eq. \eqref{eq:gamma}. Of course, anticorrelations change the low frequency behavior as discussed in the last section.

In the high frequency limit, the integral over $v$ scales as $v_{\mathrm{max}}^{3-\tau-q}$ giving an extra factor of $v_{\mathrm{max}}^{-q}$ compared to Eq. \eqref{eq:S_KFSR2}. Thus,
\begin{align}
\begin{split}
	S_K(\omega)  
	& \sim \dot{\epsilon} L^d \omega^{-q\alpha/z} \xi^{\alpha-q \alpha} \\
	& \sim L^d \omega^{-q\alpha/z} \dot{\epsilon}^{1+(q-1)\alpha \nu/\beta} . 
	\label{eq:SK4}
\end{split}
\end{align}
Note that $S_K$ scales as $L^d$ for all $\omega$ in the FSR regime, which is consistent with the observed scaling of $\Delta K \sim L^{d/2}$ in Ref. \cite{Clemmer2021}.

In the QS regime, the scaling of $S_K$ as $L^d$ required $q=1$.
If $q=1$, then the high-frequency limit of Eq. \eqref{eq:SK4} becomes identical to that for the QS regime in Eq. \eqref{kscale}.
Moreover the noise power is simply proportional to $\dot{\epsilon}$.
This is consistent with the rate dependence of $S_K (\omega)/\dot{\epsilon}$ for small systems in Fig. \ref{fig:ratedep}.
That figure shows that $S_K (\omega)/\dot{\epsilon}$ remains the same at high frequencies as $\dot{\epsilon}$ increases from the QS to FSR regime.
The response is only changed at $\omega < \omega_{\mathrm{min}}(\dot{\epsilon}) \sim \xi^{-z}$, which increases with rising rate.

\begin{figure}
\begin{center}
	\includegraphics[width=0.45\textwidth]{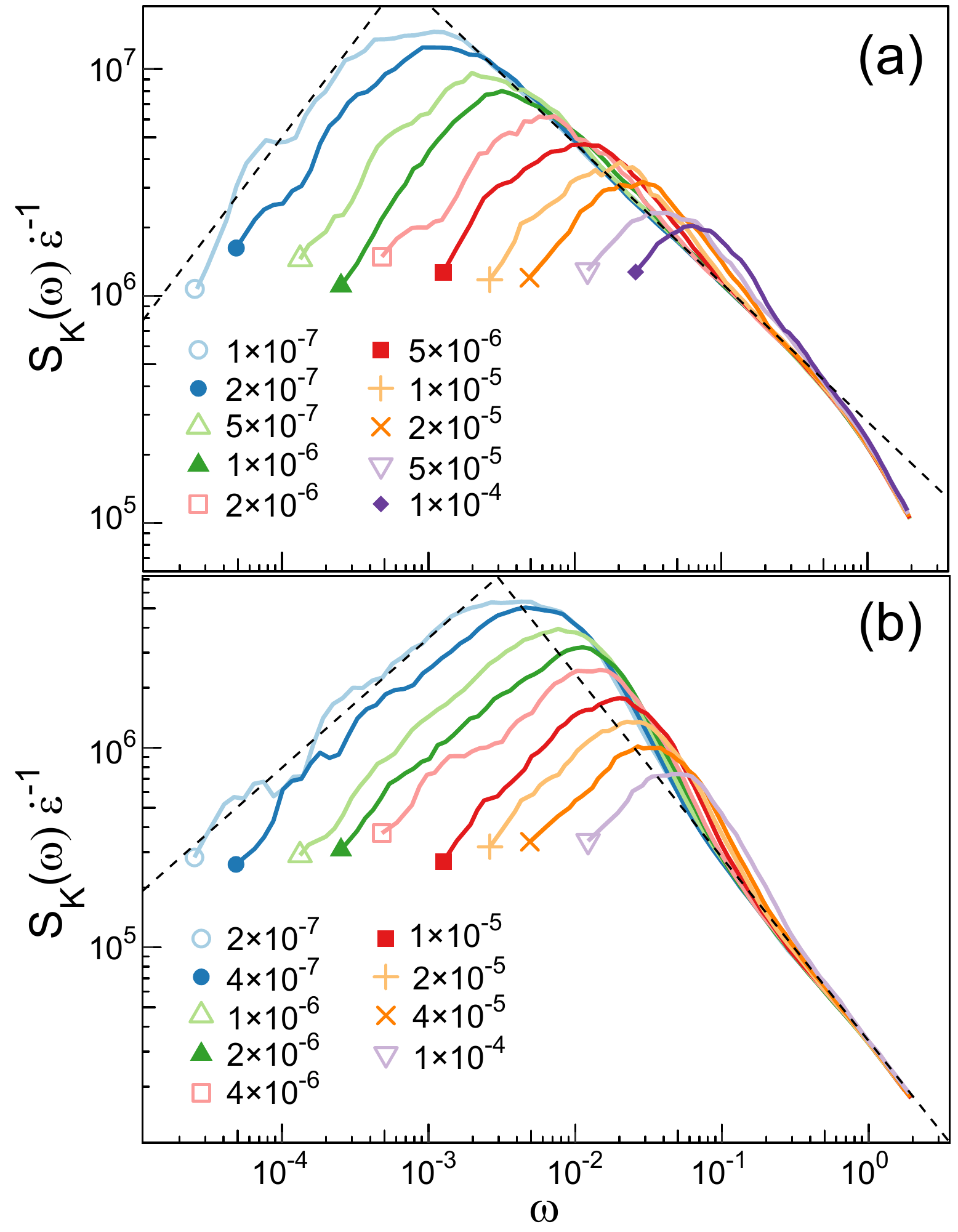}
	\caption{The power spectrum normalized by $\dot{\epsilon}$ as a function of $\omega$ for a system of size (a) $L = 1753$ in 2D and (b) $81$ in 3D at the indicated strain rates. Data at lower rates extend to lower frequencies because longer times were needed to reach comparable strain rates.  Dashed lines indicate power laws with slopes of $\eta = 0.85$ and $-q\alpha/z = -0.61$ in 2D and $\eta = 0.65$ and $-q \alpha/z = -0.92$ in 3D.
	} 
	\label{fig:LargeL}
	\end{center}
\end{figure}

Figure \ref{fig:LargeL} shows $S_K (\omega)/\dot{\epsilon}$ for large 2D and 3D systems.
Again as expected for $q=1$, the high frequency behavior is independent of rate. The main change with rate is that the high frequency power-law behavior is only seen above a cutoff frequency that rises with increasing rate.
One might expect that this steady reduction in $S_K/\dot{\epsilon}$ would lead to a reduction in $\Delta K/\dot{\epsilon}$ 
that is not evident in results for $\Delta K$ in Ref. \cite{Clemmer2021}.
However, as discussed in deriving Eq. \eqref{eq:devK2}, the integral over $S_K$ that determines $\Delta K$ is dominated by frequencies of order $\omega_{\mathrm{max}}$ where $S_K$ remains unchanged.  

The dashed lines in Fig. \ref{fig:LargeL} illustrate the power-law scaling with $\omega$.
In 2D a power law with slope $-q \alpha/z$ = $-0.61\pm0.04$ fits the low rate data over about 2 decades.
As the rate increases, the peak in the noise spectrum shifts up and to the right relative to the fit.
This appears to reflect deviations from scaling as the peak approaches $\omega_\mathrm{max}$, and these deviations become even more pronounced as $\dot{\epsilon}$ rises to the higher rates not shown.
3D simulations did not access low enough frequencies to eliminate these peaks and the range of power-law scaling is limited.
We estimate $q \alpha/z = 0.9 \pm 0.1$ from the available data noting that the range of power-law scaling is too narrow to make a precise determination.

The critical behavior in the FSR regime should be dominated by a single time scale, the longest correlation time 
$T_{\mathrm{max}}\sim \xi^{z}\sim \dot{\epsilon}^{-z\nu/\beta}$.
Given Eq. \eqref{eq:SK4}, the power
spectrum should obey a scaling law
\begin{equation}
S_K(\omega) \sim L^d \dot{\epsilon}^{1- \alpha \nu/\beta} g_\omega(\omega \dot{\epsilon}^{-z \nu/\beta}) \ \ ,
\label{eq:sw_fsr}
\end{equation}
where the high frequency response is recovered if $g_\omega(x) \sim x^{-q\alpha/z}$ for $x \gg 1$.
Because the plateau is controlled by two different time scales in the QS regime, one cannot collapse results for different rates and $L$ onto a single universal curve.

\begin{figure}
\begin{center}
	\includegraphics[width=0.45\textwidth]{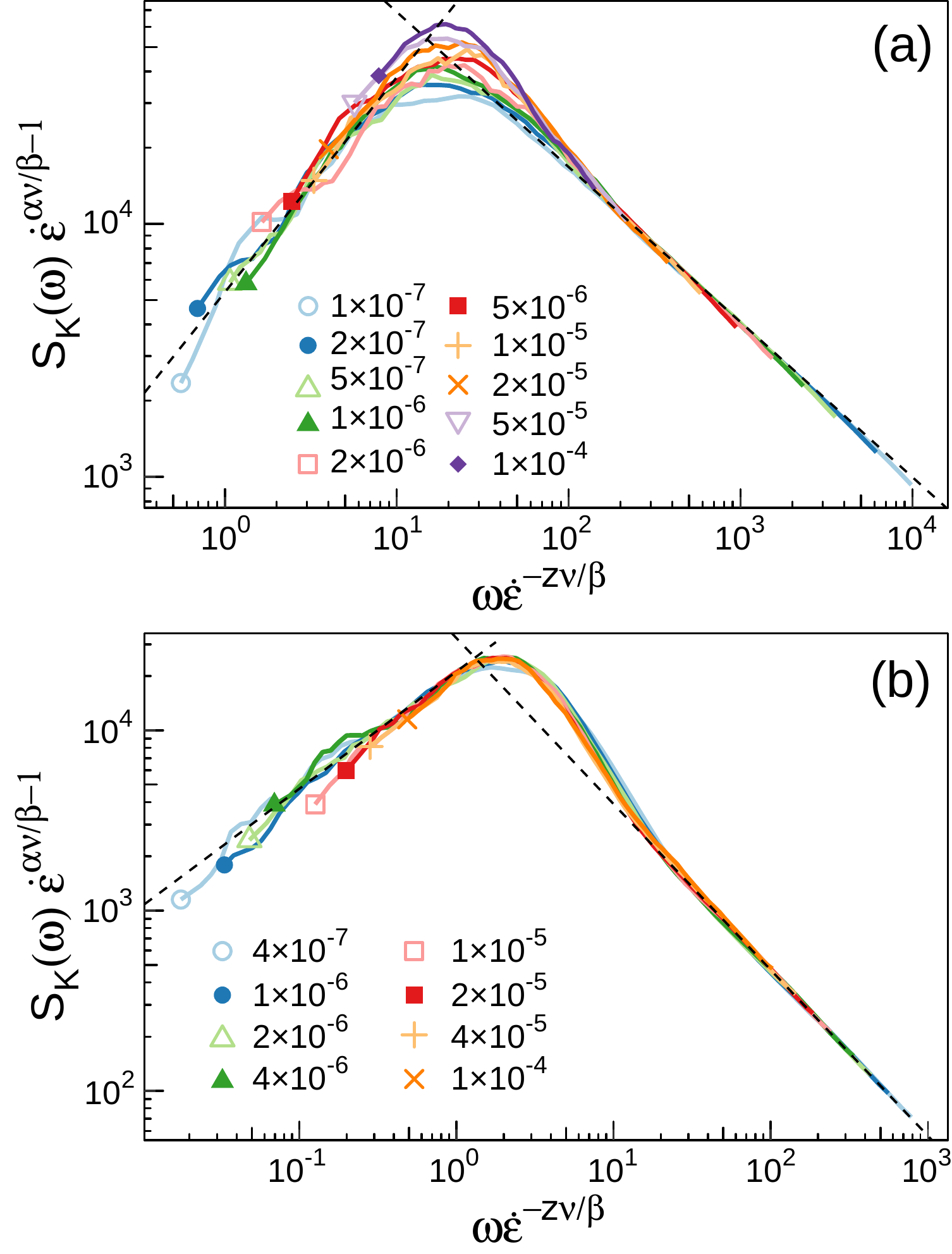}
	\caption{(a) The data in Fig. \ref{fig:LargeL}(a) is collapsed for frequencies $\omega < 0.5 \sim \omega_{\textrm{max}}$ according to Eq. \eqref{eq:sw_fsr} with exponents $z\nu/\beta = 0.62$ and $\alpha \nu/\beta = 0.38$. 
	(b) A similar scaling for the 3D data with values of $z\nu/\beta = 0.40$ and $\alpha \nu/\beta = 0.37$. 
	Dashed lines represent power laws with exponents (a) $-q \alpha/z = -0.61$ and $\eta = 0.85$ and (b) $-q \alpha/z = -0.92$ and $\eta = 0.65$.
} \label{sw_fse}
\end{center}
\end{figure}

Figure \ref{sw_fse} shows collapses of the data from Fig. \ref{fig:LargeL} using Eq. \eqref{eq:sw_fsr}. The dashed lines indicate power-law fits at small and large $\omega$.
The 2D results in Fig. \ref{sw_fse}(a) collapse at low and high $\omega$ with values of $z \nu/\beta =0.62 \pm 0.05$ and $\alpha \nu/\beta=0.39 \pm 0.04$ but show some systematic variation near the peak.
As seen in Fig. \ref{fig:LargeL}(a), the bump in $S(\omega)$ near the peak decreases as $\dot{\epsilon}$ decreases into the critical regime.
The collapse is better if the highest rates are removed. We have already noted that they appear to be outside the critical regime.
At low frequencies $S_K(\omega)$ rises with a power law of $\eta=0.85 \pm 0.10$.
The 3D results collapse well with values of $z \nu/\beta = 0.39 \pm 0.03$ and $\alpha \nu/\beta = 0.37 \pm 0.03$ in Fig. \ref{sw_fse}(b). 
There is again a characteristic bump above the high frequency power law near the peak. 
However, the fact that it shifts with the peak means that the true power-law scaling only starts at higher scaled frequencies. 
At low frequencies, we measure a power law with exponent $\eta=0.65 \pm 0.10$.
These measurements are all consistent with best estimates of exponents in Table \ref{table:yield_exponents}.

\section{Crossover From the QS to FSR Regime }
\label{sec:yield_s(w)_size}

To examine changes with rate for different $L$ we narrow our focus to the scaling of characteristic values of the power spectrum. 
First we consider the maximum value of the power spectrum, $S_\mathrm{max}(L, \dot{\epsilon})$.
From Sec. \ref{sec:yield_s(w)}, the maximum is proportional to $L^{d+\alpha} \dot{\epsilon}$ in the QS regime (Fig. \ref{fig:QSscale} and Eq. \eqref{eq:Sk_Plateau_qs}).
In contrast, in the FSR regime Eq. \eqref{eq:sw_fsr} implies $S_\mathrm{max}(L,\dot{\epsilon}) \sim \dot{\epsilon}^{1- \alpha \nu/\beta} L^{d}$.
The transition between these scaling regimes will occur at the onset of finite-size effects.  

\begin{figure}
\begin{center}
	\includegraphics[width=0.45\textwidth]{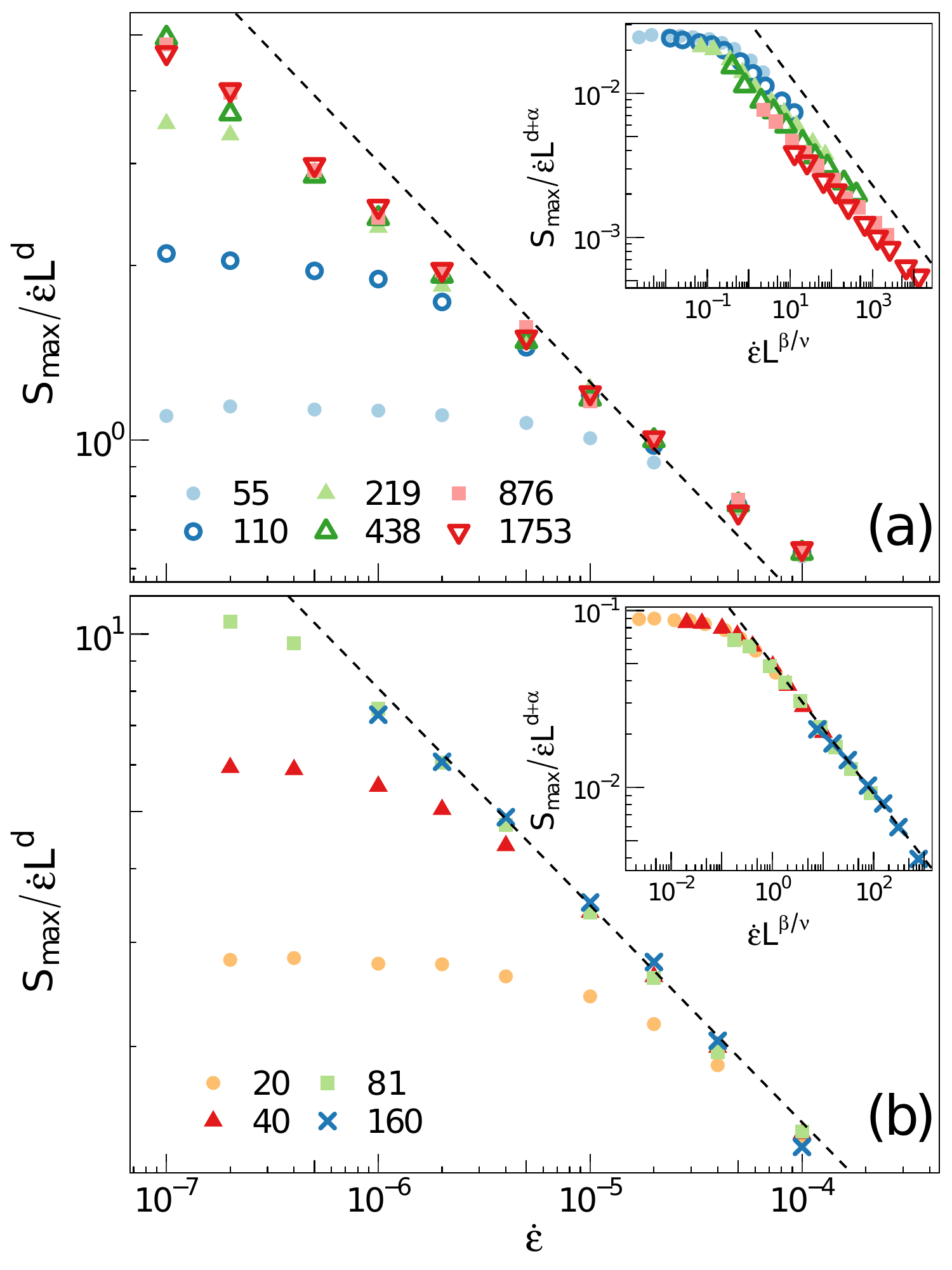}
	\caption{The maximum value of the power spectrum normalized by $\dot{\epsilon} L^{d}$ as a function of strain rate for the indicated values of $L$ in (a) 2D and (b) 3D. The data in (a) and (b) are collapsed in the insets by scaling $S_\mathrm{max}$ and $\dot{\epsilon}$ by powers of $L$ according to Eq. \eqref{smaxscale}.
	Data is scaled using exponents of $\beta/\nu = 2.5$ and $3.1$ and $\alpha = 0.95$ and $1.15$ in 2D and 3D, respectively. Dashed lines represent power laws with an exponent $- \alpha \nu/\beta =-0.38$ for 2D and -0.37 for 3D.
	} \label{sw_max}
	\end{center}
\end{figure}

In Fig. \ref{sw_max} $S_\mathrm{max}/( \dot{\epsilon} L^d)$ is plotted as a function of rate for different system sizes 
At high rates, $S_\mathrm{max}/( \dot{\epsilon} L^d)$ is independent of system size and is found to increase as a power of decreasing rate. In 3D, the data is consistent with a value of $\alpha \nu/\beta = 0.37$ derived from the exponents listed in Table \ref{table:yield_exponents}. In 2D, there is a deviation from the predicted exponent of $\alpha \nu/\beta = 0.38$. This difference is likely attributed to the evolution of the universal scaling function seen at high rates in Fig. \ref{sw_fse}. In that figure the peak height changes about a factor of 2 which is larger than the deviation from the dashed line in Fig. \ref{sw_max}(a). The deviations from scaling are less noticeable if the results are not normalized by $\dot{\epsilon}$.

At lower rates, $S_\mathrm{max}$ transitions to the expected plateau.
The crossover occurs at a lower strain rate for larger system sizes.
To capture this transition, we construct a finite-size scaling relation.
We assume there are only two relevant length scales in the problem, $L$ and $\xi$, such that the scaling will depend on the dimensionless ratio $L/\xi$ yielding the relation
\begin{equation}
	S_\mathrm{max}(L, \dot{\epsilon}) \sim \dot{\epsilon} L^{d + \alpha } f_\mathrm{max}(L^{\beta/\nu} \dot{\epsilon}) \ \ ,
\label{smaxscale}
\end{equation}
where $f_\mathrm{max}(x)$ is a new universal scaling function. To match the predicted scaling in the QS and FSR regimes, $f_\mathrm{max}(x)$ goes to a constant for $x \ll 1$ and $x^{- \alpha \nu/\beta}$ for $x \gg 1$. 
In the insets of Fig. \ref{sw_max}, curves of $S_\mathrm{max}$ versus $\dot{\epsilon}$ are rescaled according to this relation using the exponents listed in Table \ref{table:yield_exponents}. In both 2D and 3D data, curves collapse in the QS regime. However, at high rates the 2D data has some spread. As argued before, this deviation is likely due to the evolution in the universal scaling function in Fig. \ref{sw_fse}(a).

Another prominent feature of $S_K(\omega)$ is the frequency which represents the timescale of the largest avalanches, $\omega_\mathrm{min}(E_\mathrm{max})$. In the FSR regime, $E_\mathrm{max}$ is set by the correlation length and $\omega_\mathrm{min} \sim \xi^{-z}$. In the QS regime, $E_\mathrm{max}$ is limited by system size and $\omega_\mathrm{min} \sim L^{-z}$. The crossover between these two regimes occurs at the onset of finite-size effects where $\xi \sim L$. 

As seen in previous sections, $\omega_\mathrm{min}$ designates the low-frequency cutoff of the scaling $S_K(\omega) \sim \omega^{-q \alpha/z}$ observed at large $\omega$. Therefore, to measure $\omega_\mathrm{min}$ we need to identify the frequency where $S(\omega)$ transitions to this power law. We identify $\omega_\mathrm{min}$ as the minimum frequency where $d \log S(\omega)/d \log \omega$ is greater than or equal to a cutoff of $-0.2$ in 2D and $-0.5$ in 3D.  This cutoff is chosen to be between 0 and the measured value of $-q \alpha/z$. The specific value is theoretically arbitrary and not expected to affect scaling. 
Practically, the cutoff was picked to avoid underestimating $\omega_\mathrm{min}$ by detecting random fluctuations due to uncertainty in the power spectra. Results for $L=163$ were so noisy that $\omega_\mathrm{min}$ could not be determined within a factor of 3 at intermediate rates and no data is shown.

\begin{figure}
\begin{center}
	\includegraphics[width=0.45\textwidth]{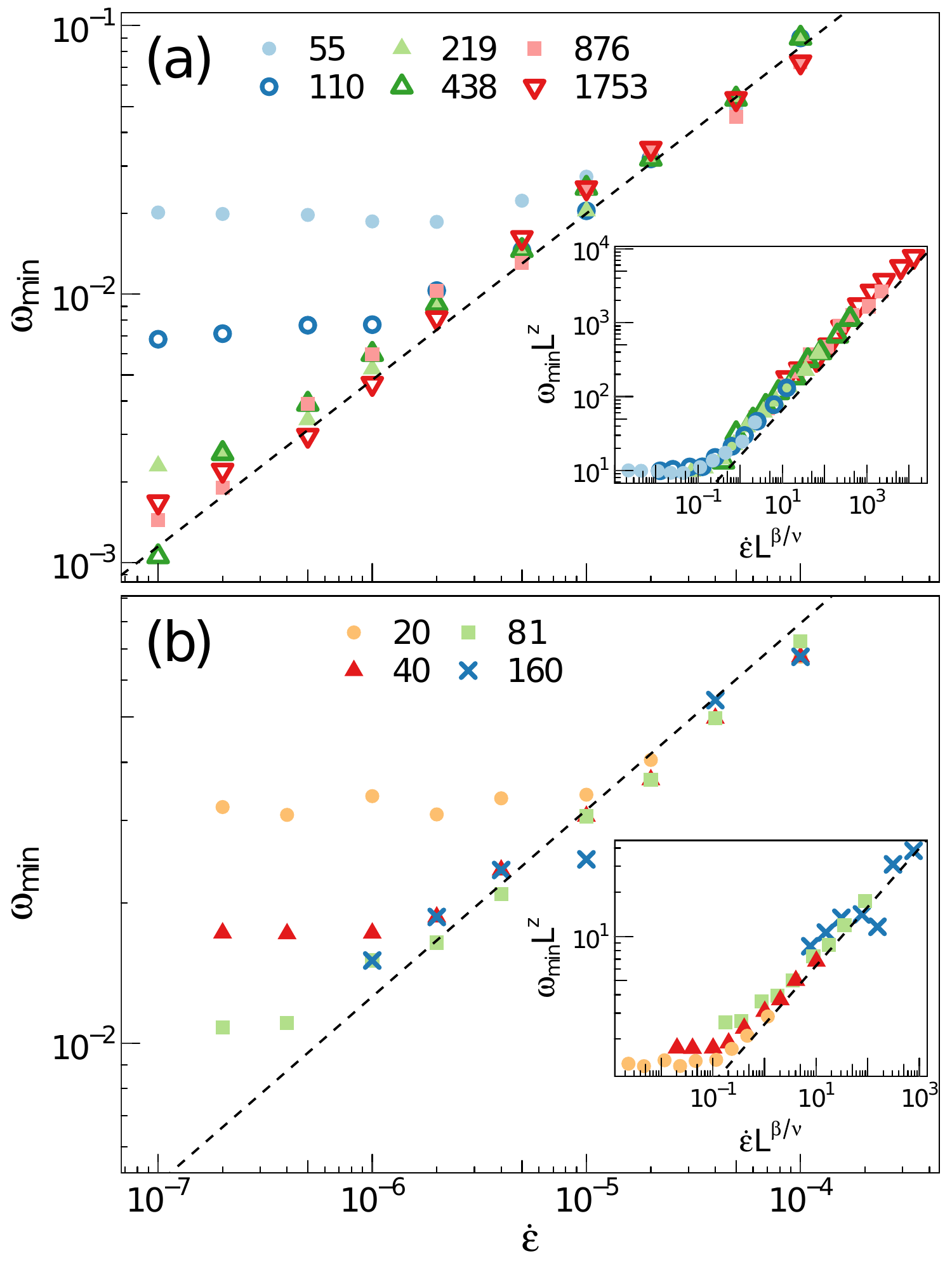}
	\caption{ $\omega_\mathrm{min}$ versus strain rate for the indicated values of $L$ in (a) 2D and (b) 3D. The data in (a) and (b) are collapsed in the insets by scaling $\omega_\mathrm{min}$ by $L^z$ and $\dot{\epsilon}$ by $L^{\beta/\nu}$ using values of $\beta/\nu = 2.5$ and $3.1$ and $z = 1.55$ and $1.25$ in 2D and 3D, respectively. Dashed lines represent power laws with an exponent $z \nu/\beta = 0.62$ in (a) and $0.4$ in (b).} \label{w_min}
	\end{center}
\end{figure}

In Fig. \ref{w_min}, $\omega_\mathrm{min}$ is plotted as a function of strain rate. At high rates, $\omega_\mathrm{min}$ depends minimally on system size and decays as a power of decreasing rate with an exponent consistent with $z \nu/\beta$ based on the exponents in Table \ref{table:yield_exponents}. As rate continues to decrease, $\omega_\mathrm{min}$ plateaus at a rate that decreases with increasing system size.

As for Fig. \ref{sw_max},
the data for $w_\mathrm{min}$
can be collapsed using a finite-size scaling relation:
\begin{equation}
\omega_\mathrm{min} \sim L^{-z} f_\mathrm{min}(L^{\beta/\nu} \dot{\epsilon}) \ \ ,
\end{equation}
where $f_\mathrm{min}(x)$ is a new scaling function. To satisfy the predicted scaling, $f_\mathrm{min}(x)$ is constant in the QS regime, $x \ll 1$, and $f_\mathrm{min}(x) \sim x^{z \nu/\beta}$ in the FSR regime, $x \gg 1$. In the insets of Fig. \ref{w_min}, this relation is used to collapse curves of $\omega_\mathrm{min}$ versus $\dot{\epsilon}$ using values of $\beta/\nu$ and $z$ in
Table \ref{table:yield_exponents}. 

\section{Summary of Results}
\label{sec:yield_summary}

In this paper, we derived a strain rate and system size dependent theory for the scaling of noise spectra in both the FSR and QS regimes.
This theory provides a scaling description for the amplitudes of the intermediate white noise regime and high frequency power-law decay regime as well as the limiting frequencies of each regime.
A similar QS theory was derived by Kuntz and Sethna \cite{Kuntz2000} who previously found the same expression for the power-law decay at high frequencies. 
As in this study, Kuntz and Sethna also argued $q = 1$ in the case of depinning magnetic domain walls. 
Our results and discussion focus on the critical behavior of the yielding transition, but the same theory could be extended to other systems, such as depinning, with power-law distributed avalanches.

Using 2D and 3D MD simulations, we measured temporal power spectra of the kinetic energy and tested the proposed theory by collapsing data across system size and strain rate.
In the QS regime, power spectra were scaled across different system sizes allowing us to accurately measure $\alpha$ and $z$ (Fig. \ref{fig:QSscale}).
In the FSR regime, spectra were scaled across strain rate using the exponents $\alpha$, $\beta/\nu$, and $z$ (Fig. \ref{sw_fse}).
Finally, key features of power spectra were scaled as a function of strain rate and system size to study the crossover between the FSR and QS regimes (Figs. \ref{sw_max} and \ref{w_min}).
Best estimates of exponents used in all figures are summarized in Table \ref{table:yield_exponents}.

\begin{center}
\begin{table}
\centering
\begin{tabular}{ |c|c|c|c|}
 \hline
  Values & 2D Estimates & 3D Estimates & Definition \\ 
 \hline
  $\alpha$ & $0.95 \pm 0.05$ & $1.15 \pm 0.05$ & $E_I \sim \ell_I^\alpha $ \\
	$\gamma$ & $1.3 \pm 0.1$ & $2.1 \pm 0.1$ & $R_{QS}(L,E) \sim L^\gamma$ \\
	$\tau$ & $1.3 \pm 0.1$ & $1.3 \pm 0.1$ & $R_{QS}(L,E) \sim E^{-\tau}$\\
  $\beta/\nu$ & $2.5$ & $3.1$ & $\xi \sim \dot{\epsilon}^{-\beta/\nu}$ \\
  $z$ & $1.55 \pm 0.05$ & $1.25 \pm 0.05$ & $T_\mathrm{I} \sim \ell_\mathrm{I}^z$ \\
  $\eta$ & $0.85 \pm 0.1$ & $0.65 \pm 0.1$ & $S_K (\omega) \sim \omega^\eta$, $\omega < \omega_\mathrm{corr}$ \\  
 \hline 
\end{tabular}
	\caption{Summary of critical exponents found here for 2D and 3D. Values of $\gamma$ and $\tau$ are quoted from Refs. \cite{Salerno2012, Salerno2013}.}
\label{table:yield_exponents}
\end{table}
\end{center}

The measured values of $\alpha$ are consistent but more accurate than results from QS MD simulations \cite{Salerno2012,Salerno2013} and the measured values of $\beta/\nu$ agree with our results in Ref. \cite{Clemmer2021}. 
In Ref. \cite{Clemmer2021}, we argued that $z \geq \alpha$ which is consistent with our exponents in this paper. 
Here we measure a dynamic exponent of $z > 1$ in both 2D and 3D as required by causality.
In studies of lattice-based EPMs with instantaneous information propagation, it was found that $z < 1$ \cite{Lin2014, Liu2016, Ferrero2019}. 
As physical restrictions on the transportation of information require $z \ge 1$, it was therefore suggested that $z = 1$ \cite{Lin2018}.  
Therefore, the dynamic critical behavior seen here is distinct from that in EPMs. Further discussion can be found in the sibling paper \cite{Clemmer2021}.

The measured power spectra reveal a novel anticorrelation regime at the lowest frequencies where the noise power grows as a power of increasing frequency with an exponent $\eta$. 
This regime extends up to a limiting frequency $\omega_\mathrm{corr}$ that corresponds to a particular value of strain (Fig. \ref{fig:ratedep}).
We propose this value of strain corresponds to the nucleation rate of the largest avalanches.
Simulations of avalanches in one dimensional sandpiles have also seen regimes of anticorrelation in power spectra at low frequencies \cite{Hwa1992, Kutnjak-Urbanc1996}. 
In these studies, a similar power-law growth was found to extend up to a limiting frequency that was associated with large discharge events that globally reduce the slope of the sandpile.
This is similar to the above proposed effect of large avalanches in the QS regime.
However, the existence of an anticorrelation regime in the FSR regime suggests this behavior does not solely depend on system-spanning events. At a finite rate, the largest avalanches may only reset a local region with a linear size of $\xi$.
The limiting frequency of the anticorrelation regime is used to theoretically bound $\beta/\nu$ by arguing the average time between large avalanches must be greater than their duration at QS strain rates.
Future scaling laws involving the exponent $\eta$ would provide valuable insight into the nature of these anticorrelations.

\begin{acknowledgments}
The authors thank Karin Dahmen and James Sethna for useful conversations. Calculations were performed at the Maryland Advanced Research Computing Center. This material is based upon work supported by the National Science Foundation under Grant No. DMR-1411144. MOR acknowledged support from the Simons Foundation.

The views and conclusions contained in this document are those of the authors and should not be interpreted as representing the official policies, either expressed or implied, of the Army Research Laboratory or the U.S. Government. The U.S. Government is authorized to reproduce and distribute reprints for Government purposes notwithstanding any copyright notation herein.

\end{acknowledgments}

\appendix

\section{Fourier Transforms and Power Spectra}
\label{sec:ke_v_s}

In this Appendix we define our normalization of the Fourier transform and power spectra and show the relation between kinetic energy and stress.
Scaling relations for the frequency dependence of the noise power are derived in the main text.

The Fourier transform of $K$ is defined as:
\begin{equation}
	\tilde{K}(\omega) \equiv \int_0^T dt e^{-i\omega t} K(t)= \sum_i \tilde{k}_i (\omega) \ \ ,
\label{eq:K_w}
\end{equation}
where the last equality sums over the Fourier transforms of individual avalanches $i$ that are assumed to be spatially or temporally separated and $T$ is the time which data was collected.
The inverse transform is
\begin{equation}
	K(t) \equiv \frac{1}{2\pi} \int d\omega e^{+i\omega t} \tilde{K}(\omega)
	\ \ .
\label{eq:K_t}
\end{equation}
The zero frequency limit is proportional to the average of $K$.
As discussed in Sec. \ref{sec:yield_individual}, the work done on the system is given by $\langle \sigma \rangle L^d \dot{\epsilon}$.
Energy balance therefore requires
\begin{equation}
	\tilde{K}(0) = \sum_i \tilde{k}_i(0) = T L^d \frac{\langle \sigma \rangle \dot{\epsilon}}{\Gamma} \ \ ,
	\label{eq:K0}
\end{equation}
using Eq. \eqref{eq:aval_e_k} to identify
\begin{equation}
	\tilde{k}_i(0) = \frac{E_i}{\Gamma }  \ \ .
	\label{eq:Ki0}
\end{equation}

The mean squared value of $K$ is given by
\begin{eqnarray}
	\langle K^2 \rangle &=& \frac{1}{T} \int_0^T dt |K(t)|^2
\nonumber	\\
	 &=& \frac{1}{T}  \iiint_0^T dt \frac{d\omega}{2\pi} \frac{d\omega^\prime}{2\pi}
	  e^{i\omega t} \tilde{K}(\omega)
	  e^{-i\omega^\prime t} \tilde{K}(-\omega^\prime) 
\nonumber \\	
 &=& \frac{1}{T}  \iint \frac{d\omega}{2\pi}
	  \frac{d\omega^\prime}{2\pi}
	  \tilde{K}(\omega)
	  \tilde{K}(-\omega^\prime)
	  \int_0^T dt
	  e^{i(\omega-\omega^\prime) t} 
\nonumber \\	
 &=& \int \frac{d\omega}{2\pi} \frac{1}{T} |\tilde{K}(\omega)|^2 \ \ .
\label{eq:meanK2v2}
\end{eqnarray}
We will define the total noise power $S_K(\omega) = \frac{1}{T}|\tilde{K}(\omega)|^2$, whose integral then gives 
$\langle K^2 \rangle$.
The mean value of $K$ is just $\langle K \rangle = \frac{1}{T} \tilde{K}(0)$.
Thus the mean squared variation in $K$ is:
\begin{equation}
\Delta K^2 \equiv	\langle |K- \langle K \rangle |^2 \rangle
	= \int \frac{d\omega}{2\pi} S_K(\omega)
	- \frac{1}{T^2} |\tilde{K}(0)|^2 \ .
	\label{eq:devK}
\end{equation}

To calculate $S_K(\omega) $ or the noise power of the stress, $S_{\sigma}(\omega)$, the time series of $K$ or $\sigma$ was divided into consecutive intervals of 10\% strain in 2D and 5\% strain in 3D. Within each interval, values of the relevant quantity were stored at times separated by the unit of time $t_0$.
A fast Fourier transform (FFT) with Hamming windowing function was used to calculate the power spectrum for each interval and the results were then averaged over all intervals. 
Each spectrum was normalized as described above so that the noise power does not depend on the duration of time over which it is calculated.
To minimize statistical fluctuations, the ensemble averaged spectrum was also averaged over intervals of angular frequency $\omega$ that have a logarithmic spacing. The curves were then further smoothed by applying a rolling mean. 
Two data points at lower and higher frequencies were included in the average. 

\begin{figure}
\begin{center}
	\includegraphics[width=0.45\textwidth]{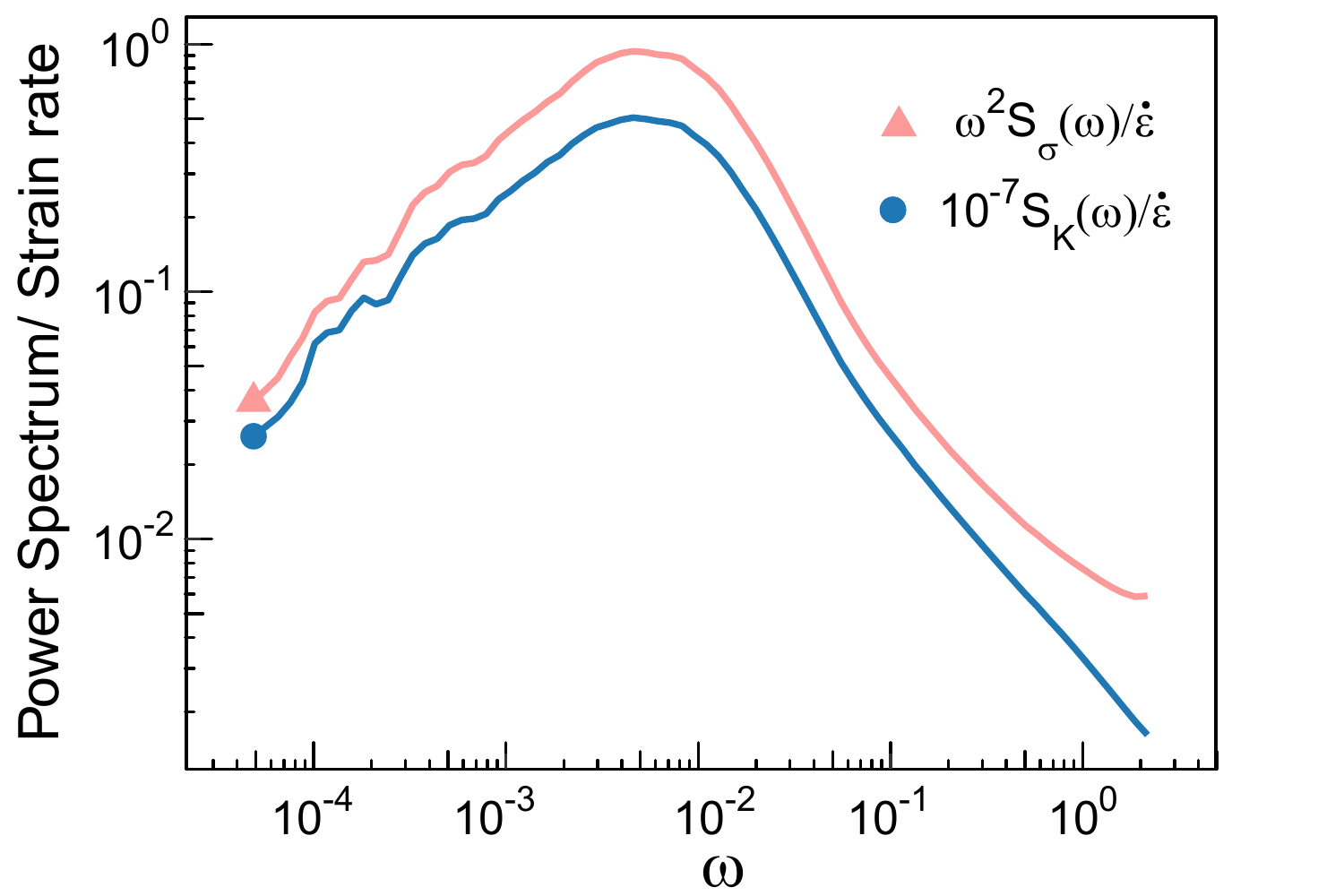}
	\caption{The power spectrum of the kinetic energy (blue, marked by a circle) and $\omega^2 S_\sigma (\omega)$ (red, marked by a triangle) as a function of $\omega$ for a system of size $L = 81$ at a rate of $4 \times 10^{-7}$. The power spectrum of the kinetic energy is shifted vertically by a factor of $10^{-7}$.
	} 
	\label{sw_kvs}
	\end{center}
\end{figure}

As noted below Eq. \eqref{eq:aval_e_k}, both $K$ and the rate of change in stress, $L^d d \sigma/dt$, are proportional to the rate of energy dissipation.
This implies $\omega^2 S_\sigma(\omega) \propto L^{-2d} S_K (\omega)$.
Figure \ref{sw_kvs} shows both quantities for a 3D system of size $L=81$ at
a rate of $4 \times 10^{-7}$ that is high enough to eliminate finite-size effects (See Fig. \ref{fig:LargeL}).
The two quantities show very similar behavior at low $\omega$.
The main difference is that artifacts related to aliasing at the highest $\omega$ lead to a slightly slower drop in $\omega^2 S_\sigma (\omega)$ than $S_K(\omega)$.
To determine the ratio of the two curves, we note that $\Gamma K$ gives the energy rate of dissipation as discussed in the beginning of Sec. \ref{sec:yield_individual}.
From Ref. \cite{Salerno2012}, the ratio of total energy dissipation in an event to the stress drop is $L^d \langle \sigma \rangle /4 \mu \sim 0.02 L^d$ where $\mu $ is the shear modulus.
Thus the ratio of $K$ to $d \sigma /dt$ should be $L^d \langle \sigma \rangle / 4 \mu \Gamma \sim 2.4 \times 10^3,$ since $\mu \sim 16$.
This gives $S_K/\omega^2 S_\sigma \approx 6 \times 10^6$, which is close to the observed ratio.
We present data for $S_{K}(\omega)$ in the main text and below, but found similar scaling collapses for $\omega^2 S_\sigma$.

\section{Simple Shear Geometry}
\label{sec:yield_shear}

\begin{figure}
\begin{center}
	\includegraphics[width=0.45\textwidth]{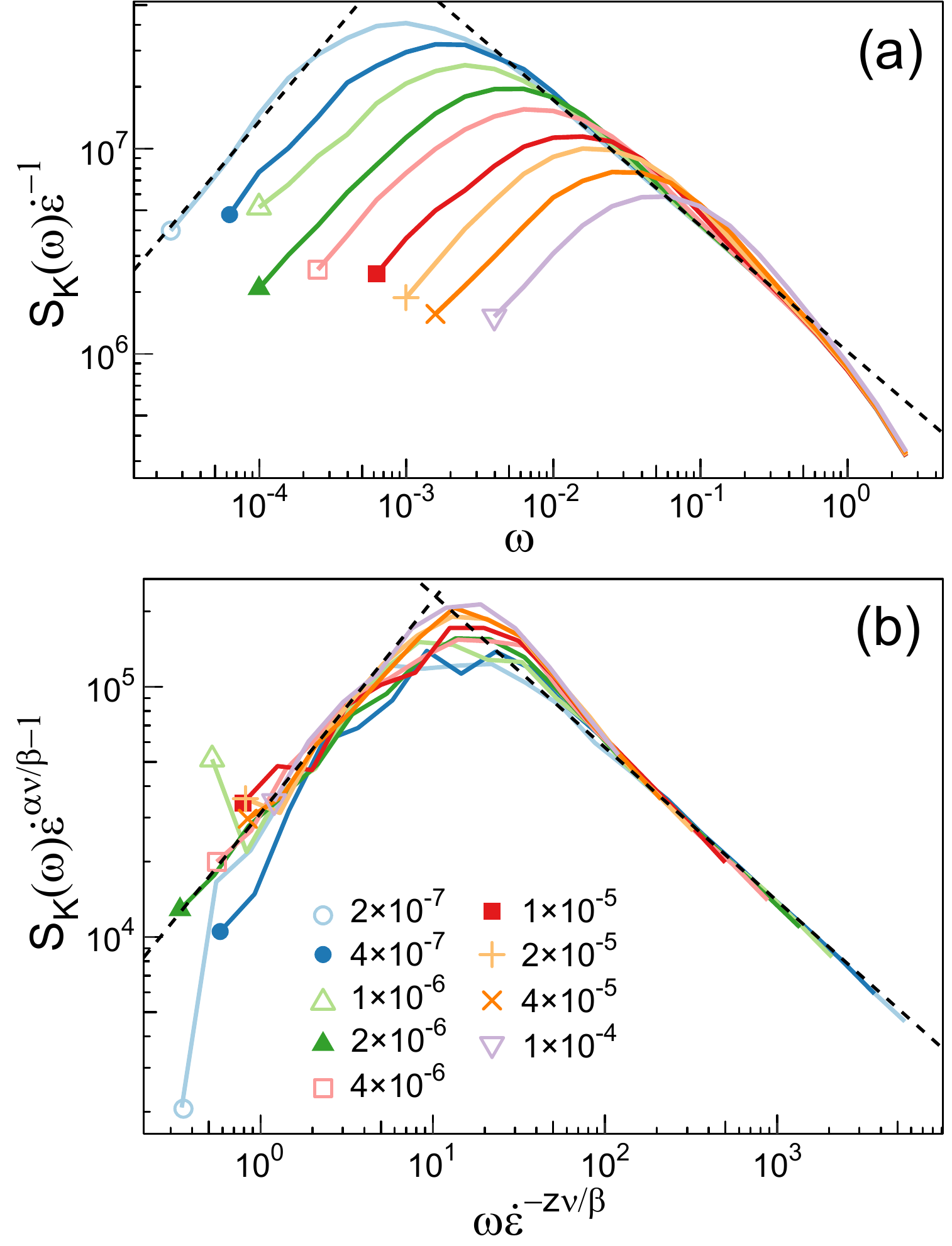}
	\caption{(a) Power spectra for a system of size $L = 876$ under simple shear at strain rates indicated in the legend. Curves are normalized by strain rate. (b) The above spectra are collapsed for $\omega < 0.5$ using the scaling relation in Eq. \eqref{eq:sw_fsr} with exponents $z \nu/\beta = 0.62$ and $\alpha \nu/\beta = 0.38$. Dashed lines in both subplots represent power laws with exponents $q \alpha/z = 0.61$ and $\eta = 0.85$} \label{fig:simple_sw}
	\end{center}
\end{figure}

The results in this paper should not only apply to pure shear or KR boundary conditions as demonstrated in this appendix where we consider power spectra from 2D simulations undergoing simple shear. 
Spectra exhibit the same scaling seen in the main text with strain rate in the FSR regime.
In Fig. \ref{fig:simple_sw}(a), $S_K(\omega)$ is plotted for simple shear of $L = 876$ systems in the FSR regime. The curves generally mirror spectra shown in
Fig. \ref{fig:LargeL} for systems using KR boundary conditions. At low $\omega$, $S(\omega)$ rises as a power of $\omega$ with an exponent consistent with $\eta = 0.85$ before peaking at a rate dependent frequency $\omega_\mathrm{corr} = \omega_\mathrm{min}$. At higher frequencies, $S(\omega)$ drops off as a power law up to a frequency $\omega_\mathrm{max} \approx 0.5$. This power-law drop-off is consistent with the previously measured exponent $q \alpha/z \sim 0.61$. 

As in Sec. \ref{sec:yield_s(w)_FSR}, the peak frequency $\omega_\mathrm{min}$ increases with rate, reflecting a reduction in the duration of the largest avalanches. 
Using the scaling relation in Eq. \eqref{eq:sw_fsr}, these curves are collapsed in Fig. \ref{fig:simple_sw}b using the exponents from Fig. \ref{sw_fse}a. 
This further supports our measured values of $\beta/\nu$ and $z$ and evidences that the noise spectra do not depend on the driving geometry.

\newpage
\bibliographystyle{apsrev4-1}

\end{document}